\documentclass[aps,prl,twocolumn,showpacs,superscriptaddress]{revtex4-1}

\usepackage{amssymb,amsmath}
\usepackage[version=4]{mhchem} %For nice and easy chemical formulas, like \ce{H20} (instead of H$_2$O)
\usepackage{graphicx}% Include figure files
\usepackage{physics} %Useful definitions of vectors, brakets etc.
  \renewcommand{\v}[1]{\vb*{#1}} %Vector (in italics, as it should be!)

\begin{document}
\title{Quantum spin ice under a $[111]$ magnetic field: \\from pyrochlore to kagom\'e}

\author{Troels Arnfred Bojesen}
\email[]{troels.bojesen@riken.jp}
\affiliation{Quantum Matter Theory Research Team, RIKEN Center for Emergent Matter Science (CEMS), Wako, Saitama 351-0198, Japan}

\author{Shigeki Onoda}
\email[]{s.onoda@riken.jp}
\affiliation{Quantum Matter Theory Research Team, RIKEN Center for Emergent Matter Science (CEMS), Wako, Saitama 351-0198, Japan}
\affiliation{Condensed Matter Theory Laboratory, RIKEN, Wako, Saitama 351-0198, Japan}

\date{\today}

\begin{abstract}
Quantum spin ice, modeled for magnetic rare-earth pyrochlores, has attracted great interest for hosting a U(1) quantum spin liquid, which involves spin-ice monopoles as gapped deconfined spinons, as well as gapless excitations analogous to photons. However, the global phase diagram under a $[111]$ magnetic field remains open. Here we uncover by means of unbiased quantum Monte-Carlo simulations that a supersolid of monopoles, showing both a superfluidity and a partial ionization, intervenes the kagom\'e spin ice and a fully ionized monopole insulator, in contrast to classical spin ice where a direct discontinuous phase transition takes place. We also show that on cooling, kagom\'e spin ice evolves towards a valence bond solid similar to what appears in the associated kagom\'e lattice model [S. V. Isakov {\it et al.}, Phys. Rev. Lett. \textbf{97}, 147202 (2006)]. Possible relevance to experiments is discussed.
\end{abstract}

%\keywords{}

\maketitle

\begin{figure}
\includegraphics[width=\columnwidth]{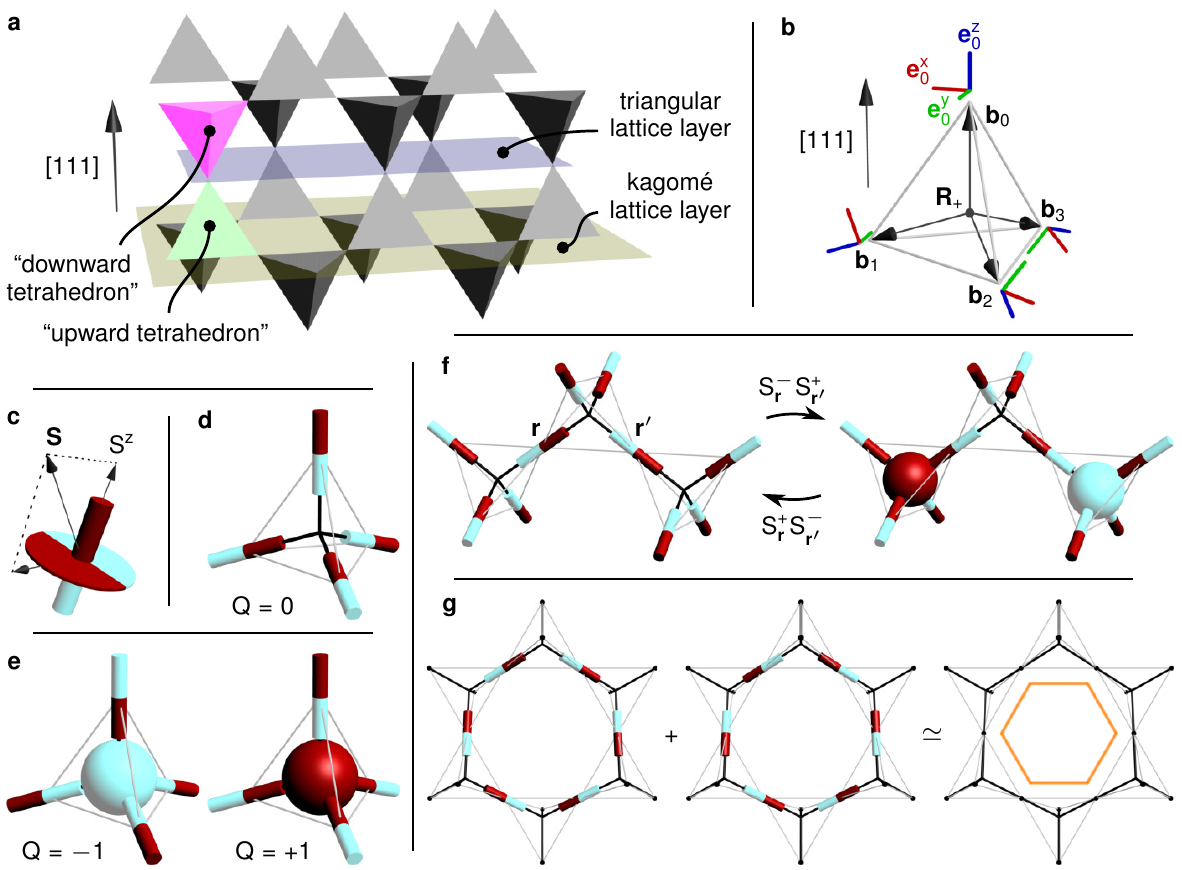}
\caption{
  \textbf{a}, The pyrochlore lattice structure, with a global $[111]$ direction and $(111)$ kagom\'e and triangular lattice layers. \textbf{b}, A pyrochlore lattice site $\v{r}=\v{R}_\pm \pm\v{b}_\mu$, with $\v{R}_\pm$ and $\v{b}_\mu$ ($\mu=0,1,2,3$) being the center of an upward/downward tetrahedron and a sublattice vector, respectively.
  \textbf{c}, A pictorial representation of a spin $\v{S}$.
  \textbf{d} and \textbf{e}, 2-in, 2-out and 3-in, 1-out (1-in, 3-out) configurations at a tetrahedron containing a $Q=0$ and $+1$ ($-1$) monopole charge, respectively.
  \textbf{f}, A spin exchange process propagates monopole charge.
  \textbf{g}, A superposition of states that can tunnel into each other by a hexagon ring exchange process.
\label{fig:basicexplanations}} 
\end{figure}

\begin{figure*}
\includegraphics[width=\textwidth]{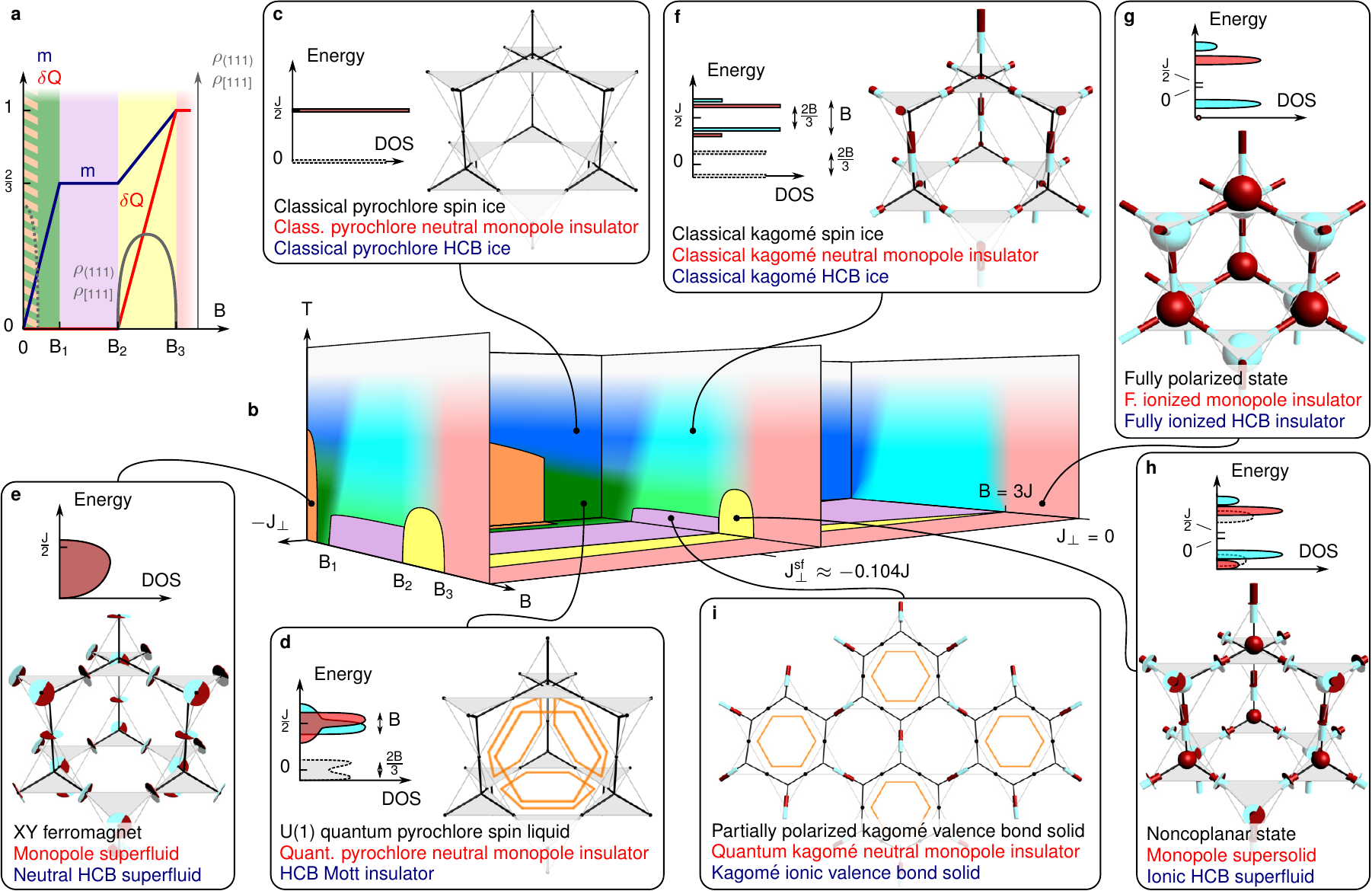}
\caption{
\textbf{a},
Schematic illustration of $m$, $\delta Q$, $\rho_{(111)}$ and $\rho_{[111]}$ as functions of $B$ in the ground state. 
\textbf{b},
The global $J_{\perp}$-$B$-$T$ phase diagram obtained by QMC simulations.
\textbf{c}--\textbf{i},
Illustrations of the excitation spectrum, as well as the average spatial profiles of spins and monopoles in each phase or regime named in the languages of pseudospins, spin-ice monopoles, and hard-core bosons (HCB)~\cite{hard-core-boson} (from top to bottom);
the classical spin ice regime (\textbf{c}), U(1) quantum spin liquid regime (\textbf{d}), monopole superfluid phase (\textbf{e}), classical kagom\'e spin ice regime (\textbf{f}), fully ionized monopole insulator with a full 3-in, 1-out spin polarization (\textbf{g}), monopole supersolid phase (\textbf{h}), and kagom\'e valence bond solid phase (\textbf{i}).
For excitation spectra, monopole charge sectors are colored based on whether $Q=-1$ (cyan), $0$ (gray), or $+1$ (red).
For graphical representations of spins and monopoles, see Fig.~\ref{fig:basicexplanations}.
In \textbf{a} and \textbf{b}, transitions and crossovers between phases and regimes are illustrated by sharp lines and gradients between different colors, respectively. In the classical limit $J_{\perp}=0$, $B_1$ decays to zero, while it is finite for $J_{\perp}\ne0$.
The valence bond solid phase ($B_1<B<B_2$) and the supersolid phase ($B_2<B_3$) have different order paramters. (See the main text.) Thus, in the current case of three dimensions, they should be separated by a first-order phase transition (at $B_2$) or a narrow coexisting phase (not shown), according to Landau theory. The orders of the transitions at $B_1$ and $B_3$ remain open due to the limitation in accessible system sizes. Photons become only two-dimensional in the kagom\'e valence bond solid, and thus are eventually gapped out by a confinement of ``dual monopoles''~\cite{polyakov:75}.
In \textbf{a}, the spin stiffnesses $\rho_{(111)}$, $\rho_{[111]}$ in the vicinity of zero field, as marked by the striped area, is zero in the quantum pyrochlore spin liquid for $J_\perp>J_\perp^{\mathrm{sf}}$ (\textbf{d}), or finite in the monopole superfluid phase for $J_\perp<J_\perp^{\mathrm{sf}}$ (\textbf{e}).
\label{fig:phasediagram}}
\end{figure*}

There exists a prototype of magnetic rare-earth pyrochlores~\cite{gardner:10} that involve a strong geometrical frustration of interactions among effective pseudospin-$1/2$ moments located at the vertices of a corner-sharing network of tetrahedra (Fig.~\ref{fig:basicexplanations}\textbf{a}). For instance, many low-temperature magnetic and thermodynamic properties of \ce{Dy2Ti2O7} and \ce{Ho2Ti2O7} \cite{harris:97,bramwell:01,moessner:06,castelnovo:08} are practically described by the nearest-neighbor antiferromagnetic Ising model,
\begin{equation}
  H_{\text{cl}}=J\sum_{\langle\v{r},\v{r}'\rangle}S^z_{\v{r}}S^z_{\v{r}'}, \quad J>0,
  \label{eq:H_z}
\end{equation}
where $\v{S}_{\v{r}}=(S^x_{\v{r}},S^y_{\v{r}},S^z_{\v{r}})$ represents a pseudospin-$\frac{1}{2}$ operator at a pyrochlore lattice site $\v{r}$ in a $C_2$-invariant set of local spin frames~\cite{onoda:11,chang:11} $(\v{e}^x_\mu,\v{e}^y_\mu,\v{e}^z_\mu)$ with the sublattice index $\mu=0,1,2,3$ (Fig.~\ref{fig:basicexplanations}\textbf{b}). This interaction forces a 2-in, 2-out spin ice rule~\cite{harris:97}: in each tetrahedron, the energy is minimized by two spins pointing inwards to and the other two outwards from the center (Fig.~\ref{fig:basicexplanations}\textbf{d}), in an analogy to proton displacements in hexagonal water ice~\cite{bernal:33}. This leaves a residual ice entropy associated with the macroscopic degeneracy of the spin-ice-rule vacuums. Creating 3-in, 1-out or 1-in, 3-out local defects -- monopoles -- with which we assign a charge $Q=+1$ or $-1$ (Fig.~\ref{fig:basicexplanations}\textbf{e}), costs half the spin-ice-rule interaction energy, $J/2$. These monopoles behave as static quasiparticles obeying an analogous Coulomb law and can only be excited thermally~\cite{%isakov:04,
moessner:06,castelnovo:08}. The average low-temperature pseudospin and monopole charge configuration, as well as its excitation spectrum, is depicted in Fig.~\ref{fig:phasediagram}\textbf{c}.

On the other hand, \ce{Yb2Ti2O7}~\cite{chang:11,ross:11}, \ce{Tb2Ti2O7}~\cite{takatsu:15}, and \ce{Pr2Zr2O7}~\cite{kimura:13} have been understood as quantum spin ice~\cite{onoda:10,onoda:11} where spin-flip exchange interactions, for instance,
\begin{equation}
  H_\perp= J_\perp\sum_{\langle {\v{r}},{\v{r}}'\rangle}
  \left( S^{x}_{\v{r}} S^{x}_{{\v{r}}'} + S^{y}_{\v{r}} S^{y}_{{\v{r}}'}\right),
  \label{eq:H_XX}
\end{equation}
become active in the background of the spin-ice-rule interaction $H_{\text{cl}}$. Spin flips are accompanied by a transfer of monopole charge (Fig.~\ref{fig:basicexplanations}\textbf{f}), so that the monopoles exhibit quantum kinematics as bosonic spinons, leading to a broadening of charge-1 exciations (Fig.~\ref{fig:phasediagram}\textbf{d}). If the spin-ice rule interaction dominates over the kinetic energy, the monopoles remain incompressible with an energy gap in their excitations. Then, a ``diamagnetic current'' and the associated ``flux'' generated by successive spin flips around closed paths (Fig.~\ref{fig:basicexplanations}\textbf{g}) may be fixed. This deconfines the monopoles and leaves gapless spin excitations described by ``photons'' in a magnetic analogue of quantum electrodynamics~\cite{hermele:04,moessner:06} (Fig.~\ref{fig:phasediagram}\textbf{d} with $B=0$). The quantum spin ice is now in what is called a U(1) quantum spin liquid state~\cite{hermele:04} which can also be viewed as a quantum pyrochlore neutral monopole insulator. This has been evidenced by quantum Monte-Carlo simulations on the minimal $H_{\text{XXZ}}=H_{\text{cl}}+H_\perp$ for $0>J_\perp>J_\perp^{\mathrm{sf}}=-0.104J$~\cite{banerjee:08,kato:15}. Conversely, if the kinetic energy dominates over the spin-ice-rule interaction, as is the case when $J_\perp<J_\perp^{\mathrm{sf}}$, the monopoles are Bose-Einstein condensed and thus confined~\cite{lee:12}, resulting in a monopole superfluid (Fig.~\ref{fig:phasediagram}\textbf{e}). Note that the superfluid density of monopoles is proportional to a transverse spin stiffness $\rho$
\footnote{{
The spin stiffness parallel and perpendicular to the $[111]$ field is given by
$
\rho_{[111]} = -\frac{1}{\beta L^3}\langle I_{[111]}^2 \rangle,
  \quad
  \rho_{(111)} = -\frac{1}{2\beta L^3}\langle I_{(111)}^2 \rangle
$
with $I^z_{[111]} = \v{e}^{z}_0 \cdot \v{I}^z$ and $\v{I}^z_{(111)} = \v{I}^z - I^z_{[111]}\v{e}^{z}_0$, where
\begin{equation*}
  \v{I}^z = i\frac{J_\perp}{2}\sum_{\langle\v{r},\v{r}'\rangle} (\v{r}-\v{r}')\left( S^{+}_{\v{r}} S^{-}_{{\v{r}}'} - S^{-}_{\v{r}} S^{+}_{{\v{r}}'}\right)
  \label{eq:I}
\end{equation*}
represents the total spin current associated with $S^z$.
This is exactly half the total monopole current $\v{I}^m$;
\begin{align*}
 \v{I}^m ={}& i\frac{J_\perp}{2}\sum_{\tau=\pm}\tau\sum_{\v{R}_\tau}
  \sum_{\mu\ne\nu}2(\v{b}_\mu-\v{b}_\nu)
  \Phi_{\v{R}_\tau+2\tau\v{b}_\mu}^\dagger \\
  &\times e^{-i(A_{\v{R}_\tau,\v{R}_\tau+2\tau\v{b}_\mu}-A_{\v{R}_\tau,\v{R}_\tau+2\tau\v{b}_\nu})}
  \Phi_{\v{R}_\tau+2\tau\v{b}_\nu} \ ,
\end{align*}
and therefore, the spin stiffness is nothing but the superfluid stiffness of monopoles.
In the above equation, similarly to Ref.~\cite{lee:12}, we have introduced a canonical conjugate pair of directed diamond-lattice link variables of an analogous ``electric field'' $E_{\v{R}_+,\v{R}_++2\v{b}_\mu}=-S^z_{\v{R}_++\v{b}_\mu}$ and ``vector potential'' $A_{\v{R},\v{R}+\v{b}_\mu}$ satisfying $[E_{\v{R}_+,\v{R}_++2\v{b}_\mu},A_{\v{R}'_+,\v{R}'_++2\v{b}_\nu}]=i\delta_{\v{R}_+,\v{R}_+'}\delta_{\mu,\nu}$, as well as spinon operators $\Phi_{\v{R}}=e^{-i\varphi_{\v{R}}}$ and $\Phi^\dagger_{\v{R}}=e^{i\varphi_{\v{R}}}$ decreasing and increasing monopole charge by 1, respectively, with $\varphi_{\v{R}}$ being a phase canonical conjugate to the monopole charge $Q_{\v{R}_\tau}=\sum_\mu E_{\v{R}_\tau,\v{R}_\tau+2\tau\v{b}_\mu}$
}}.
Hence, a finite monopole superfluid density points to an XY ferromagnet of psuedospins.

Now, of our interest is the fate of quantum spin ice against a $[111]$ magnetic field $\v{B}=B\v{e}_0^z$. The Hamiltonian reads
\begin{equation}
  H=H_{\text{XXZ}}-\v{B}\cdot\sum_{\v{R}_+}\v{m}_{\v{R}_+},
  \quad
  \v{m}_{\v{R}_+}=\sum_{\mu=0}^3\v{e}^z_\mu S^z_{\v{R}_++\v{b}_\mu}
  \label{eq:H}
\end{equation}
with the magnetization $\v{m}_{\v{R}_+}$ of the ``upward'' oriented tetrahedron centered at $\v{R}_+$ and the sublattice vector $\v{b}_\mu$ measured from $\v{R}_+$. We have assumed that only $S^z_{\v{r}}$ couples to the magnetic field, as is the case for non-Kramers ions \ce{Pr^3+} and \ce{Tb^{3+}}~\cite{onoda:09,onoda:10}, e.g. in \ce{Pr2Ir2O7}~\cite{machida:10}, \ce{Pr2Zr2O7}~\cite{kimura:13}, \ce{Pr2Hf2O7}~\cite{sibille:16}, and \ce{Tb2Ti2O7}~\cite{mirebeau:07,takatsu:15}, where $S^x_{\v{r}}$ and $S^y_{\v{r}}$ correspond to electric quadrupoles~\cite{onoda:09,onoda:10,takatsu:15}. In the classical limit~\cite{moessner:01,moessner:03,isakov:04}, $J_\perp=0$, $B$ produces two successive transitions at zero temperature $T=0$. (See the $J_\perp=0$ plane of Fig.~\ref{fig:phasediagram}\textbf{b}.) First, an infinitesimally small $B$ forces spins in the $(111)$ triangular-lattice layers (Fig.~\ref{fig:basicexplanations}\textbf{a}) to point in the field direction, i.e, $\langle S^z_{\v{R}_++\v{b}_0}\rangle=1/2$, while the remaining spins in the $(111)$ kagom\'e-lattice layer take a 2-in, 1-out configuration in each upward tetrahedron, leading to $\sum_{\mu=1}^3\langle S^z_{\mathrm{R}_++\v{b}_\mu}\rangle=-1/2$ (Fig.~\ref{fig:phasediagram}\textbf{f}) and a magnetization plateau at $m=|\langle\v{m}_{\v{R}_+}\rangle|=2/3$. The extensive degeneracy of spin ice partially remains within each kagom\'e layer, and hence the state is called kagom\'{e} spin ice~\cite{cornelius:01,fukazawa:02,sakakibara:03}. At $B=3J$, there occurs an abrupt spin-flip transition from 2-in, 2-out to the 3-in, 1-out fully polarized state with $m=1$. Accordingly, the monopole charge disproportionation $\delta Q \equiv \langle Q_{\v{R}_+}\rangle = -\langle Q_{\v{R}_-}\rangle$ jumps from 0 to 1 (Fig.~\ref{fig:phasediagram}\textbf{g}), leading to a fully ionized monopole insulator. ($\v{R}_-$ is the center of a ``downward'' oriented tetrahedron), This transition has been dubbed a monopole crystallization~\cite{brooks-bartlett:14}. Now systematic theoretical understandings in the quantum case are called for~\cite{carrasquilla:15}.

\begin{figure}
\includegraphics[width=\columnwidth]{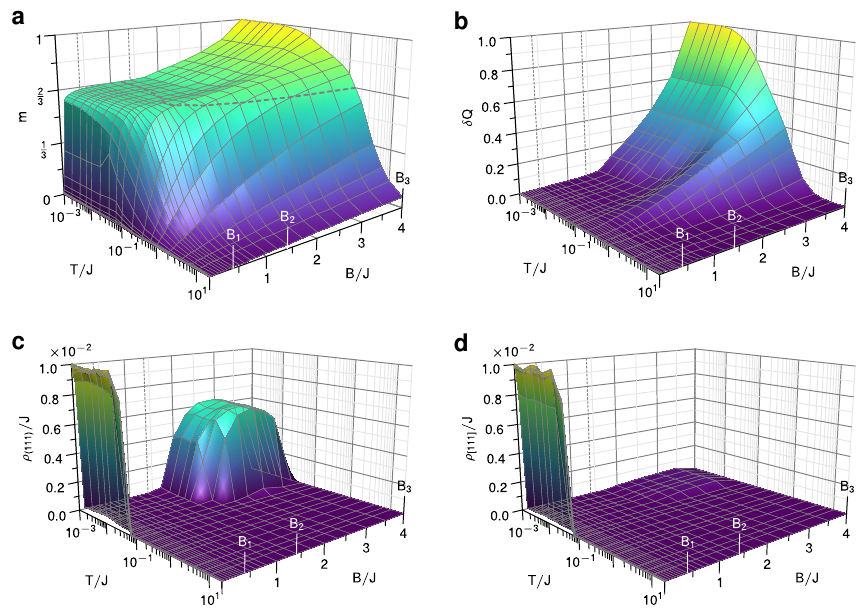}
\caption{
\textbf{a}, The $[111]$ magnetization per spin.
\textbf{b}, The monopole charge disproportionation per tetrahedron.
\textbf{c} and \textbf{d}, The transverse spin stiffness normal and parallel to the $[111]$ magnetic field.
For all plots, $J_{\perp}=-0.15J$ and $L=10$. Each mesh vertex represents one QMC data point.
Statistical errorbars are invisibly small, except in the low-temperature region of the monopole superfluid phase in \textbf{c} and \textbf{d}, where they are at most $3\times10^{-4}$. Careful numerical annealing was employed on cooling to combat severe freezing problems.
\label{fig:thermodynamics}}
\end{figure}

We perform continuous imaginary-time world-line quantum Monte-Carlo simulations \cite{kawashima:04,kato:07,kato:15} on the minimal quantum spin ice model given by Eq.~\eqref{eq:H} with $J_\perp < 0$. Figure~\ref{fig:thermodynamics} presents results on $m$, $\delta Q$ and two components of the spin stiffness, $\rho_{(111)}$ and $\rho_{[111]}$, being normal and parallel to the field, for the particular case of $J_{\perp}=-0.15J<J_{\perp}^{\mathrm{sf}}$, in which the zero-field ground state is a monopole superfluid. Increasing $B$ up to $\sim0.1J$ at the lowest temperature, $m$ arises from 0 with a finite slope, i.e., a finite $[111]$ magnetic susceptibility $\chi_{[111]}$. Both $\rho_{(111)}$ and $\rho_{[111]}$ steeply decay to zero, indicating that the monopole superfluid dies out quickly. Further increasing $B$ up to $B_1\sim0.4J$, $m$ increases to $2/3$ without any apparent singularity. In these low-field ranges, monopoles are prevented from living on a long-time scale by the spin-ice-rule interaction, so $\delta Q=0$. Similar behaviors of continuously increasing $m$, i.e., $\delta Q=0$ and $\rho_{(111)}=\rho_{[111]}=0$, appear from $B=0$ when $J_\perp=-0.09J>J_\perp^{\mathrm{sf}}$. 

Using the previous estimate of the velocity of the photons, $1.49(4)(a|J_\perp|^3/J^2\hbar)$, for $B=0$~\cite{kato:15}, we find a ground-state value of $\chi_{[111]}\sim240$ for $J_\perp=-0.09J$. This indicates that the Curie law displayed by $\chi_{[111]}$ in the classical spin ice regime~\cite{isakov:04} is cut off by the photons that lift the macroscopic degeneracy of the spin-ice manifold. It is therefore natural to assert that the phase out of the monopole superfluid around $B=0$ is adiabatically connected to the case with $J_\perp>J_\perp^{\mathrm{sf}}$ at $B=0$, and hence it is a neutral monopole insulator, namely, a U(1) quantum spin liquid. From $B_1$ to $B_2\sim1.4J$, $m$ is pinned to the $2/3$ plateau where the spin-ice-rule constraint remains to be satisified on a long-time scale, i.e. $\delta Q=0$, as in kagom\'e spin ice. In these field ranges, a nonzero $\delta Q$ appears only with moderately large thermal excitations at around $T\sim0.2$-$0.3J$ where a Schottky peak appears in the specific heat for $B=0$~\cite{kato:15}. 

Increasing $B$ above $B_2$, $m$ resumes growing from $2/3$ and simultaneously, $\delta Q$, $\rho_{(111)}$, and $\rho_{[111]}$ start increasing from zero. This evidences that the positive/negative monopole charge sectors on ``upward''/``downward'' oriented tetrahedra become softened and Bose condensed. This establishes a supersolid~\cite{andreev:69,chester:70,leggett:70} of monopoles showing a partial charge disproportionation $0<\delta Q<1$ of monopoles and a long-range transverse spin order (Fig.~\ref{fig:phasediagram}\textbf{h}). Note that $m$ is not an order parameter at finite magnetic field. The supersolid is distinguished from the superfluid at zero field by having a finite monopole charge disproportionation. The spin stiffness is strongly anisotropic with $\rho_{(111)}$ being an order of magnitude larger than $\rho_{[111]}$, indicating that the transverse spin order is triggered by correlations within the kagom\'e layers.

A further increase in $B$ drives a phase transition at $B_3 \sim 4J$ to the fully ionized monopole insulator characterized by $\delta Q=1$ and $\rho_{(111)}=\rho_{[111]}=0$. Reflecting that the monopole supersolid emerges because of a kinetic energy gain of monopoles, this phase shrinks with decreasing $|J_\perp|$ and is absent in classical spin ice systems. The lowest-temperature results are schematically summarized in Fig.~\ref{fig:phasediagram}\textbf{a}, and the global phase digram in Fig.~\ref{fig:phasediagram}\textbf{b}.

\begin{figure}
\includegraphics[width=\columnwidth]{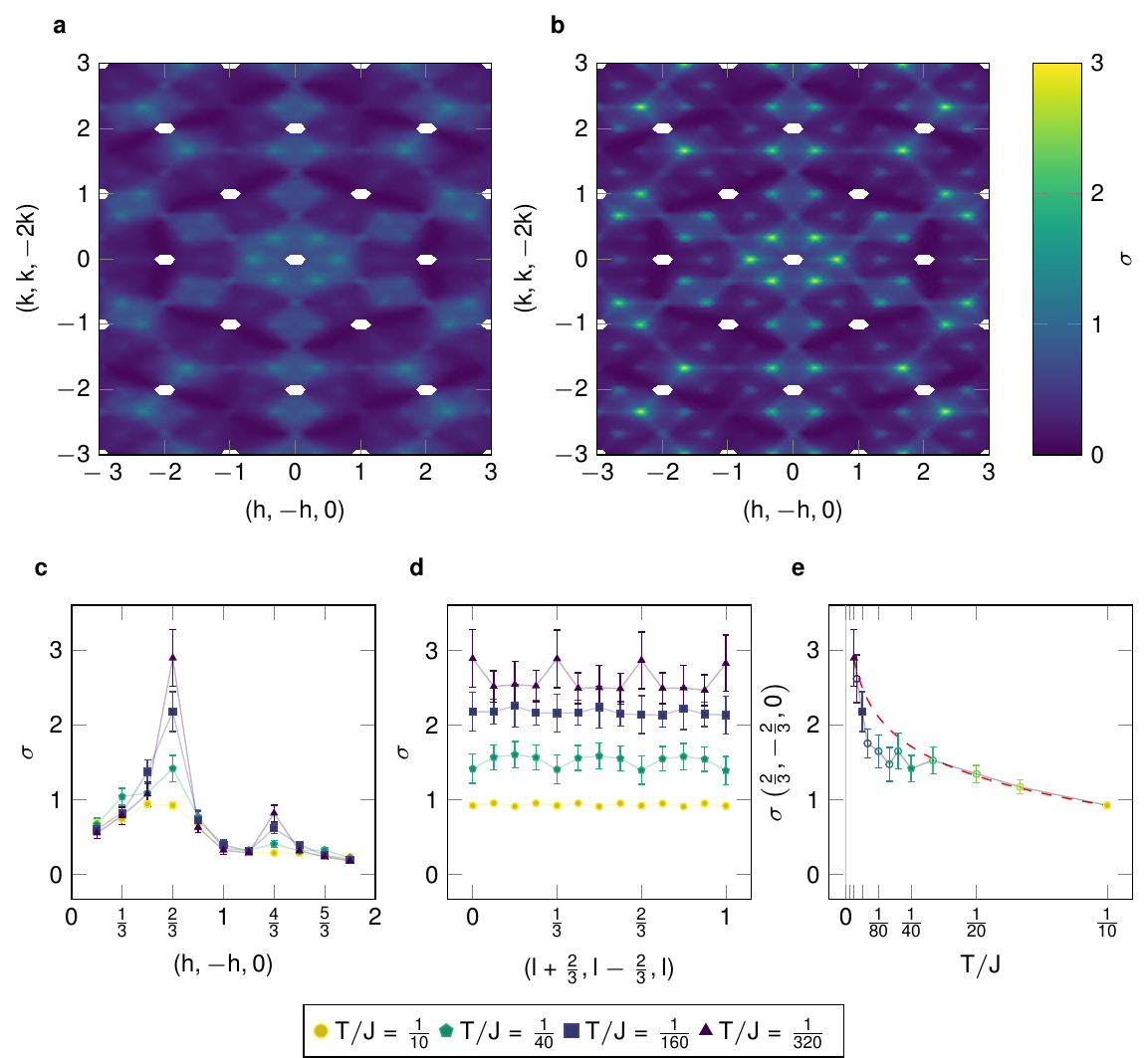}
\caption{
Quantum Monte-Carlo results~\cite{simulation_details} on equal-time (energy-integrated) neutron-scattering cross-sections
$\sigma(\v{q})=\sum_{\mu,\nu=0}^3\langle S^z_{\mu,\v{q}}S^z_{\nu,-\v{q}}\rangle\left(\v{e}^z_\mu\cdot\v{e}^z_\nu-\frac{(\v{q}\cdot\v{e}^z_\mu)(\v{q}\cdot\v{e}^z_\nu)}{q^2}\right)$ with $S^z_{\mu,\v{q}}=\frac{1}{L^{3/2}}\sum_{\v{R}}e^{-i\v{q}\cdot(\v{R}+\v{b}_\mu)}S^z_{\v{R}+\v{b}_\mu}$. 
\textbf{a} and \textbf{b}, Spatial profiles in the $(h+k,-h+k,-2k)$ plane in the classical kagom\'e spin ice regime, $(J_{\perp},B,T) = (-0.15,1.3,1/20)J$, and in a short-range ordered state towards the kagom\'e valence bond solid~\cite{isakov:06}, $(J_{\perp},B,T) = (-0.15,1.3,1/320)J$, respectively. 
\textbf{c} and \textbf{d}, Temperature profiles along lines going through the position $\v{q}_\text{sl}=(2\pi/a)(\frac{2}{3},-\frac{2}{3},0)$ of the superlattice Bragg spot observed in the associated 2D quantum kagom\'e spin ice model~\cite{isakov:06}.
\textbf{e}, Temperature dependence of the peak intensity at $\v{q}_\text{sl}$. The red dashed curve is a fit to the $\log T$ dependence expected for the nearest-neighbor classical kagom\'e spin ice~\cite{moessner:03}. For all the panels, the labels for $\v{q}$ are given in reciprocal lattice units of $2\pi/a$. The Bragg peaks due to the polarized triangular lattice spins have been subtracted.
\label{fig:neutronscattering}}
\end{figure}

To understand the low-temperature properties in the kagom\'e spin ice plateau regime, we compute the spatial profiles of energy-integrated diffuse neutron-scattering cross-sections. Figure~\ref{fig:neutronscattering}\textbf{a} shows the profile in the classical kagom\'e spin ice regime at $T=J/20$ and $B=1.3J$. There appear short-range correlations associated with a broad peak at $\v{q}_\text{sl} = \frac{2\pi}{a}(\frac{2}{3},-\frac{2}{3},0)$ and symmetry-related points, in addition to a broadened pinch-point singularity at $\v{q}=\frac{2\pi}{a}(\frac{2}{3},\frac{2}{3},-\frac{4}{3})$ and symmetry-related points, with $a$ being the cubic lattice constant. The pattern clearly matches the experimental observation in \ce{Dy2Ti2O7}~\cite{tabata:06}. On cooling down to $T=J/320$, short-range correlations apparently develops around the superlattice point $\v{q}_\text{sl}$ and symmetry-related points on the (111) plane (Fig.~\ref{fig:neutronscattering}\textbf{b}). This correlation around $\v{q}_\text{sl}$ is found to be anisotropic. Along the cut $\frac{2\pi}{a}(h,-h,0)$ within the (111) plane, the peak sharpens on cooling (Fig.~\ref{fig:neutronscattering}\textbf{c}), while along an out-of-plane direction $\frac{2\pi}{a}(l+\frac{2}{3},l-\frac{2}{3},l)$, the intensity remains to be flat within errorbars (Fig.~\ref{fig:neutronscattering}\textbf{d}). Besides, while the peak intensity at $\v{q}_\text{sl}$ exhibits a logarithmic increase on cooling as in classical kagom\'e spin ice~\cite{moessner:03}, it starts being saturated at $T\sim J/30$, indicating that a lifting of the extensive degeneracy of the classical kagom\'e spin ice manifold is visible on this energy scale. Then, it restarts increasing more rapidly below $T\sim0.01J$ (Fig.~\ref{fig:neutronscattering}\textbf{e}). It is likely that the ground state has a two-dimensional long-range order enlarging the unit cell by $\sqrt{3}\times\sqrt{3}$ as in the single-layer quantum kagom\'e spin ice model~\cite{isakov:06} (Fig.~\ref{fig:phasediagram}\textbf{i}). At present, however, it is difficult to reliably collect lower temperature data on our pyrochlore model with quantum Monte-Carlo simulations. It remains open whether this valence bond solid forms a two-dimensional or three-dimensional pattern.

So far, there has been no concrete experimental evidence of the U(1) quantum spin liquid in candidate quantum spin ice materials at zero magnetic field. Nevertheless, praseodymium pyrochlores remain to be promising candidates, since diffuse neutron-scattering patterns in \ce{Pr2Zr2O7} at zero magentic field~\cite{kimura:13} are consistent with the previous numerical simulation on the same model indicating the emegent photon modes~\cite{kato:15}. Also, a step in the magnetization curve has already been observed in \ce{Pr2Ir2O7}~\cite{machida:10} most likely as a precursor to the $2/3$ magnetization plateau. A careful annealing under a $[111]$ magnetic field might lead to the quantum kagom\'e valence bond solid. Then, it will also be possible to observe the monopole supersolid by increasing the field above the plateau and measuring the electric quadrupole moments with polarized neutron scattering experiments. The monopole supersolid phase, if observed, is a manifestation of a quantumness in spin ice and of monopoles.

\begin{acknowledgments}
The work was partially supported by Grants-in-Aid for Scientific Research under Grant No. 24740253 and No. 15H03692 from Japan Society for the promotion of Science and under Grant No. 15H01025 from the Ministry of Education, Culture, Sports, and Technology of Japan  and by the RIKEN iTHES project.
Numerical computationss were performed by using the HOKUSAI-Great Wave supercomputing system at RIKEN.
\end{acknowledgments}

\bibliography{references}

\begin{thebibliography}{39}
\bibitem [{har()}]{hard-core-boson}%
  \BibitemOpen
  \href@noop {} {}\bibinfo {note} {Note that the model Eq.~\eqref{eq:H} and the
  current results can also be interpreted in terms of hard-bosons, as proposed
  for a possible relevance to cold atom systems~\cite{isakov:04,banerjee:08}.
  The $[111]$ magentic field plays a role of ionic potentials differentiating
  kagom\'{e} and triangular lattice sites.}\BibitemShut {Stop}%
\bibitem [{sim()}]{simulation_details}%
  \BibitemOpen
  \href@noop {} {}\bibinfo {note} {$L=12$ and 16 independent parallel
  simulations were performed with the following series of cooling steps: $1/T =
  0.01,0.02,0.04,$ $\ldots,0.32,0.64,1,2,3,$
  $\ldots,99,100,102,104,106,\ldots$. At each temperature, $2\times 10^5$ worm
  loops were generated for equilibration of the state inherited from the
  previous higher temperature. Then, measurements were performed during
  $3\times 10^6$ loop updates. Statistical errors where computed by bootstrap
  analysis.}\BibitemShut {Stop}%
\end{thebibliography}

\end{document}